\documentclass[12pt,a4paper]{article}
\usepackage{longtable}
\usepackage{hyperref}
\usepackage[english]{babel}
\usepackage{setspace}
\usepackage[dvips]{graphics,color}      
\usepackage{verbatim}

\def\R2Lurl#1#2{\mbox{\href{#1}{\tt #2}}}

\title{The posterior-Viterbi: a new decoding algorithm for hidden
Markov models}

\author{Piero Fariselli*, Pier Luigi Martelli, and Rita Casadio \\
\\
Department of Biology, University of Bologna\\
via Irnerio 42, 40126 Bologna, Italy \\
Tel: +39-051-2091280 \quad Fax: +39-051-242576 \\
\\
e-mails: piero.fariselli@unibo.it,
\\
gigi@biocomp.unibo.it,
\\
casadio@alma.unibo.it
\\
\bigskip
$\ast$ To whom correspondence should be addressed
\bigskip
\bigskip
\bigskip
}

\date{}

\begin{document}

\maketitle

\newpage

\section*{ABSTRACT}

\textbf{Background:} Hidden Markov models (HMM) are powerful machine 
learning tools successfully applied to problems of computational 
Molecular Biology. In a predictive task, the HMM is endowed with 
a \textit{decoding algorithm} in order to assign the most probable 
state path, and in turn the class labeling, to an unknown sequence. 
The Viterbi and the posterior decoding algorithms are the most common. 
The former is very efficient when one path dominates, while the 
latter, even though does not guarantee to preserve the automaton 
grammar, is more effective when several concurring paths have 
similar probabilities. A third good alternative is 1-best, which 
was shown to perform equal or better than Viterbi.\\
\textbf{Results:} In this paper we introduce the posterior-Viterbi 
(PV) a new decoding which combines the posterior and Viterbi 
algorithms. PV is a two step process: first the posterior probability 
of each state is computed and then the best posterior allowed 
path through the model is evaluated by a Viterbi algorithm.\\
\textbf{Conclusions:} We show that PV decoding performs better than 
other algorithms first on toy models and then on the computational 
biological problem of the prediction of the topology of beta-barrel 
membrane proteins. \\
\textbf{Contacts}: piero.fariselli@unibo.it

\section*{Background}

Machine learning approaches have been shown to be very profitable 
in the field of computational Molecular Biology \cite{BaldiBrunak2001}. 
Among them  hidden Markov models (HMMs) have been proven to be especially successful 
when in the problem at hand regular grammar-like structures can 
be detected \cite{BaldiBrunak2001,Durbinetal.1998}. 
HMMs were developed for alignments \cite{Kroghetal.1994, Baldietal.1994}, 
pattern detection \cite{Mamitsuka1998,Batemanetal.2002} 
and also for predictions, as in the 
case of the topology of all-alpha and all-beta membrane proteins 
\cite{TusnadySimon1998,Kroghetal.2001, Martellietal.2002, 
Martellietal.2003, Liuetal.2003, ViklundElofsson2004, Bagosetal.2004, Bigelowetal.2004}. 

When HMMs are implemented for predicting a given feature, a \textit{decoding} 
algorithm is needed. With decoding we refer to the assignment 
of a path through the HMM states (which is the best under \textit{a 
suitable measure}) given an observed sequence \textit{O}. In this way, 
we can also assign a class label to each sequence element of 
the emitting state \cite{BaldiBrunak2001, Durbinetal.1998}. 
More generally, as stated in \cite{Krogh1997}, 
the \textit{decoding is the prediction of the labels of an unknown 
path}. Labeling is routinely the only relevant biological property 
associated to the observed sequence; the states themselves may 
not represent a significant piece of information, since they basically define 
the automaton grammar.
 
The most famous decoding procedure is the Viterbi algorithm, 
which finds the most probable allowed path through the HMM model. 
Viterbi decoding is particularly effective when there is a \textit{single 
best path} among others much less probable. When several paths 
have similar probabilities, the posterior decoding or the 1-best 
algorithms are more convenient \cite{Krogh1997}. 
The posterior decoding assigns 
the state path on the basis of the posterior probability, although 
the selected path might be not allowed. For this reason, in order 
to recast the automaton constraints, a post-processing algorithm 
was applied to the posterior decoding \cite{Farisellietal.2003}.

In this paper we address the problem of preserving the automaton 
grammar and concomitantly exploiting the posterior probabilities, 
without the need of the post-processing algorithm \cite{Farisellietal.2003,
Martellietal.2002}. 
Prompted by this, we design a new decoding algorithm, the \textit{posterior-Viterbi 
decoding} (PV), which \textit{preserves the automaton grammars and 
at the same time exploits the posterior pobabilities}. We show 
that PV performs better than the other algorithms when we test 
it on toy models and on the problem of the prediction of the topology 
of beta-barrel membrane proteins.

\section*{Methods}

\subsubsection*{The hidden Markov model definitions}

For sake of clarity and compactness, in what follows we make 
use of explicit \textit{BEGIN} and \textit{END} states and we do not treat 
the case of the \textit{silent} (\textit{null}) states. Their inclusion 
in the algorithms is only a technical matter and can be done 
following the prescriptions indicated in \cite{BaldiBrunak2001,Durbinetal.1998}.

An observed sequence of length \textit{L} is indicated as \textit{O} (=$O_1...O_L$)
both for a single-symbol-sequence (as in the standard HMMs) or 
for a vector-sequence as described before \cite{Martellietal.2002}. 
\textit{label}(\textit{s}) indicates the label associated to the 
state \textit{s}, while $\Lambda$ (=$\Lambda_i,\dots \Lambda_L$) 
is the list of the labels associated to each sequence position \textit{i} 
obtained after the application of a decoding algorithm. Depending 
on the problem at hand, the labels may identify transmembrane 
regions, loops, secondary structures of proteins, coding/non 
coding regions, intergenic regions, etc. A HMM consisting of \textit{N} 
states is therefore defined by three probability distributions

{\bf Starting probabilities:}
\begin{eqnarray}
a_{BEGIN, k}= P(k|BEGIN)
\end{eqnarray}

{\bf Transition probabilities:}
\begin{eqnarray}
a_{k,s}= P(k|s)
\end{eqnarray}

{\bf Emission probabilities:}
\begin{eqnarray}
e_{k}(O_{i})= P(O_{i}|k) 
\end{eqnarray}

\noindent
The forward probability is 
\begin{equation}
f_{k}(i) = P(O_{1},O_{2}\dots O_{i},\pi_{i}=k)
\end{equation}
which is the probability of having emitted the first partial 
sequence up to \textit{i} ending at state \textit{k}. 

\noindent
The backward probability is: 
\begin{equation}
b_{k}(i) = P(O_{i+1},\dots O_{L-1},O_{L}|\pi_{i}=k)
\end{equation}
which is the probability of having emitted the sequence starting 
from the last element back to the (\textit{i}+1)th element, given 
that we end at position \textit{i} in state \textit{k}. The probability 
of emitting the whole sequence can be computed using either forward 
or backward according to: 
\begin{equation}
P(O|M)=f_{END}(L+1)=b_{BEGIN}(0)
\end{equation}
Forward and backward are also necessary for updating of the HMM 
parameters, using the Baum-Welch algorithm \cite{BaldiBrunak2001, Durbinetal.1998}. 
Alternative a gradient-based training algorithm can be applied \cite{BaldiBrunak2001, Krogh1997}.

\subsubsection*{Viterbi decoding}

Viterbi decoding finds the path ($\pi$) through the model 
which has the maximal probability with respect to all the others 
\cite{BaldiBrunak2001, Durbinetal.1998}. 
This means that we look for path  which is
\begin{equation}
\pi^{v}= argmax_{\{\pi\}}P(\pi|O,M)
\label{eq:vit1}
\end{equation}
where $O$(=$O_{1},\dots O_{L}$) is the observed sequence of length $L$ and $M$ 
is the trained HMM model. 
Since the $P(O|M)$ is independent of a particular path $\pi$, Equation \ref{eq:vit1}
is equivalent to
\begin{equation}
\pi^{v}= argmax_{\{\pi\}}P(\pi,O|M)
\end{equation}
$P(\pi,O|M)$ can be easily computed as
\begin{equation}
P(\pi,O|M)=\prod_{i=1}^{L}a_{\pi(i-1),\pi(i)}e_{\pi(i)}(O_i)\cdot a_{\pi(L),END}
\end{equation}
where by construction $\pi(0)$ is always the $BEGIN$ state.

Defining $v_{k}(i)$ as the probability of the most likely path ending  
in state $k$ at position $i$, 
and $p_i(k)$ as the trace-back pointer, $\pi^{v}$ can 
be obtained running the following dynamic programming called 
Viterbi decoding

\begin{itemize}
\item {\bf Initialization}
\begin{eqnarray*}
v_{BEGIN}(0)=1 & v_{k}(0)=0 & for \quad k \neq BEGIN
\end{eqnarray*}

\item {\bf Recursion}
\begin{eqnarray*}
v_{k}(i) = [ \max_{\{s\}}(v_{s}(i-1)a_{s,k})] e_{k}(O_{i}) \\
p_i(k)= argmax_{\{s\}} v_{s}(i-1)a_{s,k}
\end{eqnarray*}

\item {\bf Termination}
\begin{eqnarray*}
P(O,\pi^v |M) = \max_{\{s\}}[v_s(L)a_{s,END}] \\
\pi^v_L=argmax_{\{s\}}[v_s(L)a_{s,END}]
\end{eqnarray*}

\item {\bf Traceback}
\begin{eqnarray*}
\pi^v_{i-1}=p_i(\pi^v_{i}) & for \quad i=L \dots 1
\end{eqnarray*}
\item {\bf Label assignment}
\begin{eqnarray*}
\Lambda_i=label(\pi^v_i) & for \quad i=1 \dots L
\end{eqnarray*}
\end{itemize}

\subsection*{1-best decoding}

The 1-best labeling algorithm described here is the Krogh's previously 
described variant of the N-best decoding \cite{Krogh1997}. Since there is 
no exact algorithm for finding the most probable labeling, 1-best 
is an approximate algorithm which usually achieves good results 
in solving this task \cite{Krogh1997}. Differently from Viterbi, the 1-best 
algorithm ends when the most probable labeling is computed, so 
that no trace-back is needed.

For sake of clarity, here we present a redundant description, 
in which we define $H_i$ as the set of all labeling hypothesis surviving 
as 1-best for each state $s$ up to sequence position $i$. 
In the worst case the number of distinct labeling-hypothesis is equal 
to the number of states. $h_i^s$ is the current partial labeling hypothesis 
associated to the state $s$ from the beginning to the \textit{i}-th 
sequence position. In general several states may share the same 
labeling hypothesis. 
Finally, we use $\oplus$ as the {\em string concatenation operator}, so that
'AAAA'$\oplus$'B'$=$'AAAAB'.
1-best algorithm can then described as
\begin{itemize}
\item {\bf Initialization}
\begin{eqnarray*}
v_{BEGIN}(0)=1 & v_{k}(0)=0 \quad for \quad k \neq BEGIN \\
v_k(1)=a_{BEGIN,k}\cdot e_k(O_1) & H_1=\{label(k) : a_{BEGIN,k}\neq 0 \} \\
H_{i}=\emptyset  & for \quad i=2, \dots L
\end{eqnarray*}

\item {\bf Recursion}
\begin{eqnarray*}
v_{k}(i+1) = & \max_{h\in H_i}[\sum_s v_{s}(i)\cdot \delta(h^s_i,h)\cdot a_{s,k})] e_{k}(O_{i}) \\
h^k_{i+1}= & argmax_{h\in H_i} [\sum_s v_{s}(i)\cdot \delta(h^s_i,h)\cdot a_{s,k})]\oplus label(k) \\
H_{i+1} \leftarrow & H_{i+1} \quad \bigcup \quad \{h^k_{i+1}\} &
\end{eqnarray*}

\item {\bf Termination}
\begin{eqnarray*}
\Lambda=argmax_{h\in H_L} \sum_s v_s(L)\delta(h_L^s,h)a_{s,END}
\end{eqnarray*}
\end{itemize}
where we use the Kronecker's delta $\delta(a,b)$ (which is 1 when $a=b$, 0 otherwise).
With 1-best decoding we do not need keeping backtrace
matrix since $\Lambda$ is computed during the forward steps.

\subsubsection*{Posterior decoding}

The $posterior$ decoding finds the path which maximizes the
product of the $posterior$ probability of the states
\cite{BaldiBrunak2001,Durbinetal.1998}. Using the usual notation
for forward ($f_{k}(i)$) and backward ($b_{k}(i)$)
we have
\begin{equation}
P(\pi_i=k|O,M)=f_k(i)b_k(i)/P(O|M)
\label{eq:pos}
\end{equation}
The path $\pi^p$ which maximizes the posterior probability
is then computed as
\begin{eqnarray}
\pi^p_i=argmax_{\{s\}} P(\pi_i=s|O,M) & for \quad i=1 \dots L
\end{eqnarray}
The corresponding label assignment is 
\begin{eqnarray}
\Lambda_i=label(\pi^p_i) & for \quad i=1 \dots L
\end{eqnarray}
If we have more than one state sharing the same label, labeling 
can be improved by summing over the states that share the same 
label (\textit{posterior sum}). In this way we can have a path through 
the model which maximizes the posterior probability of being 
in a state with \textit{label} \textit{\ensuremath{\lambda}} 
when emitting the observed sequence element , or more formally: 
\begin{eqnarray}
\Lambda_i=argmax_{\{\lambda\}}\sum_{label(s)=\lambda} P(\pi_i=s|O,M) & for \quad i=1\dots L
\end{eqnarray}

The posterior-decoding drawback is that the state path sequences $\pi^p$ 
or \ensuremath{\Lambda} may be not allowed paths. However, this decoding 
can perform better than Viterbi, when more than one high probable 
path exits \cite{BaldiBrunak2001, Durbinetal.1998}. 
In this case a post-processing algorithm that 
recast the original topological constraints is recommended \cite{Farisellietal.2003}. 

In the sequel, if not differently indicated, with the term \textit{posterior} 
we mean the posterior sum.

\subsubsection*{Posterior-Viterbi decoding}

Posterior-Viterbi decoding is based on the combination of the 
Viterbi and posterior algorithms. After having computed the \textit{posterior} 
probabilities we use a Viterbi algorithm to find the best \textit{allowed 
posterior} path through the model. A related idea, specific for pairwise 
alignments was previously introduced to improve the sequence 
alignment accuracy \cite{HolmesDurbin1998}.
 
In the PV algorithm, the basic idea is to compute the path  $\pi^{PV}$
\begin{eqnarray}
\pi^{PV}=argmax_{\{\pi \in A_p\}}\prod_{i=1}^L P(\pi_i|O,M)
\end{eqnarray}
where $A_p$  is the set of the allowed paths through the model, and  
$P(\pi_i|O,M)$ is the \textit{posterior} probability of the state assigned by the 
path \ensuremath{\pi} at position \textit{i} (as computed in Eq. \ref{eq:pos}). 

Defining a function $\delta^*(s,t)$  that is 1 if  $s \rightarrow t$ 
is an allowed transition of the model \textit{M}, 0 otherwise,  
$v_k(i)$ as the probability of the most 
probable \textit{allowed-posterior} path ending at state \textit{k} having 
observed the partial $O_1,\dots O_i$ and $p_i$ as the trace-back pointer, 
we can compute the best path $\pi^{PV}$ using the Viterbi algorithm 
\begin{itemize}
\item {\bf Initialization}
\begin{eqnarray*}
v_{BEGIN}(0)=1 & v_{k}(0)=0 & for \quad k \neq BEGIN
\end{eqnarray*}

\item {\bf Recursion}
\begin{eqnarray*}
v_{k}(i) = \max_{\{s\}}[v_{s}(i-1)\delta^*(s,k)] P(\pi_i=k|O,M) \\
p_i(k)= argmax_{\{s\}}[v_{s}(i-1)\delta^*(s,k)] 
\end{eqnarray*}

\item {\bf Termination}
\begin{eqnarray*}
P(\pi^{PV} |M,O) = max_{s}[v_s(L)\delta^*(s,END)] \\
\pi^{PV}_L=argmax_{\{s\}}[v_s(L)\delta^*(s,END)]
\end{eqnarray*}

\item {\bf Traceback}
\begin{eqnarray*}
\pi^{PV}_{i-1}=p_i(\pi^{PV}_{i}) & for \quad i=L \dots 1
\end{eqnarray*}
\item {\bf Label assignment}
\begin{eqnarray*}
\Lambda_i=label(\pi^{PV}_i) & for \quad i=1 \dots L
\end{eqnarray*}
\end{itemize}

\subsubsection*{Datasets }

\begin{figure}[htbp]
\begin{center}
{\par \scalebox{0.4}{{\includegraphics{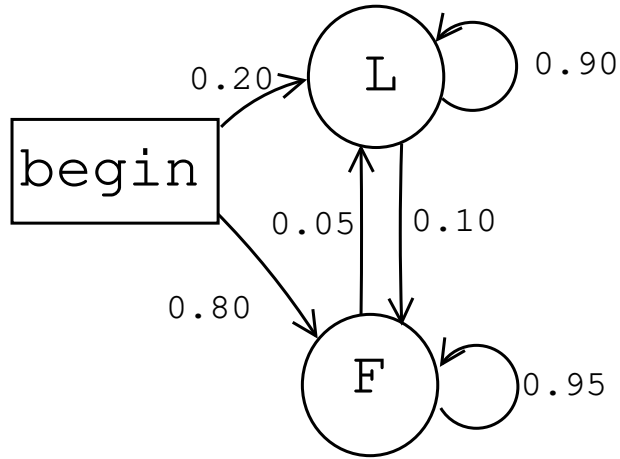}}}\par}
\caption{Occasionally dishonest casino (Model LF). The emission probabilities of the
fair state (F) are $1/6$ for each possible outcome, while in the loaded die 
the emission probabilities are $1/2$ for the '1' and $1/10$ for the other faces.}
\label{fig:lf}
\end{center}
\end{figure}

\begin{figure}[htbp]
\begin{center}
{\par \scalebox{0.4}{{\includegraphics{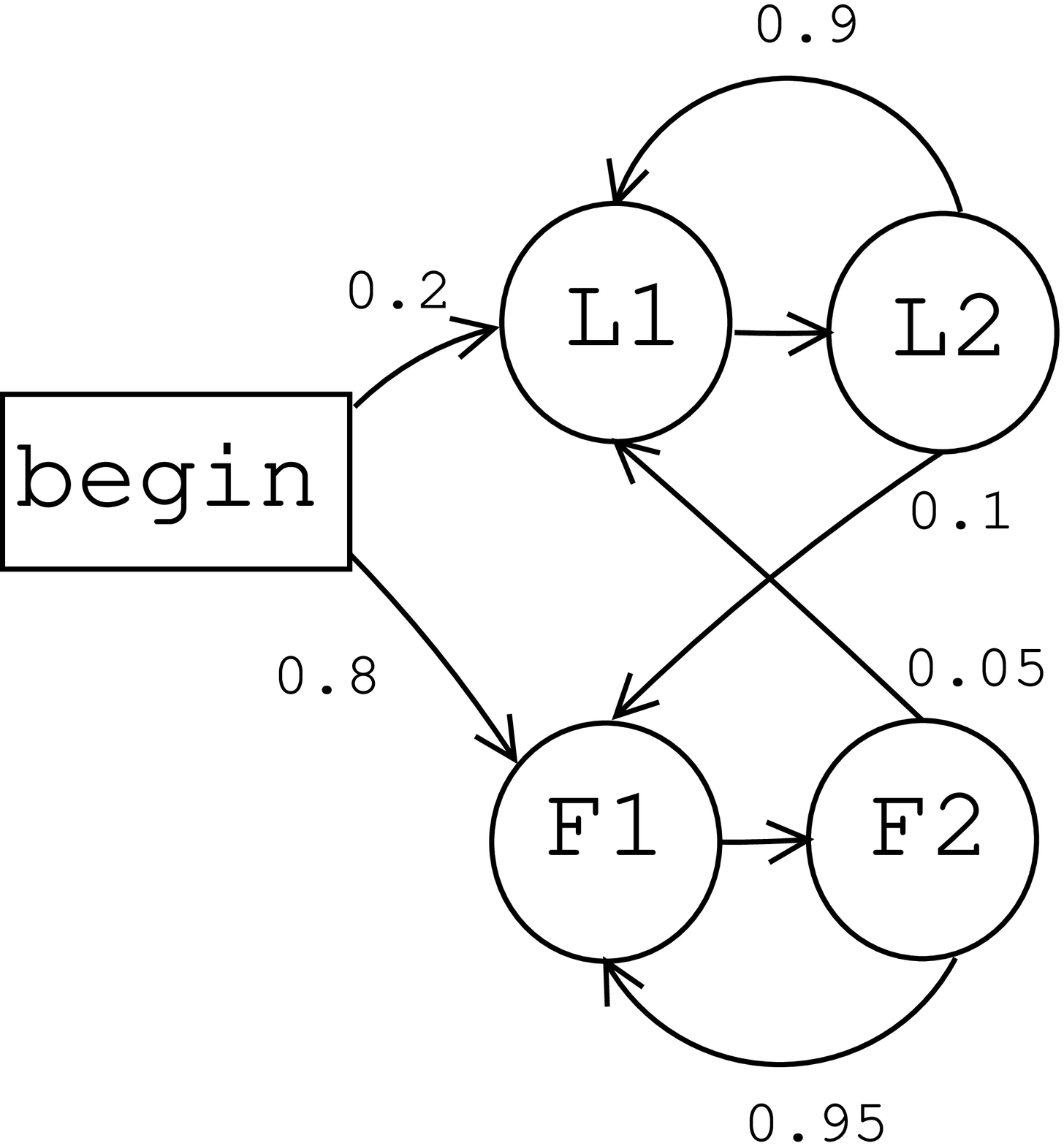}}}\par}
\caption{Occasionally dishonest casino (Model L2F2). For the emission probabilities see
Figure \ref{fig:lf}.}
\label{fig:l2f2}
\end{center}
\end{figure}

\begin{figure}[htbp]
\begin{center}
{\par \scalebox{0.4}{{\includegraphics{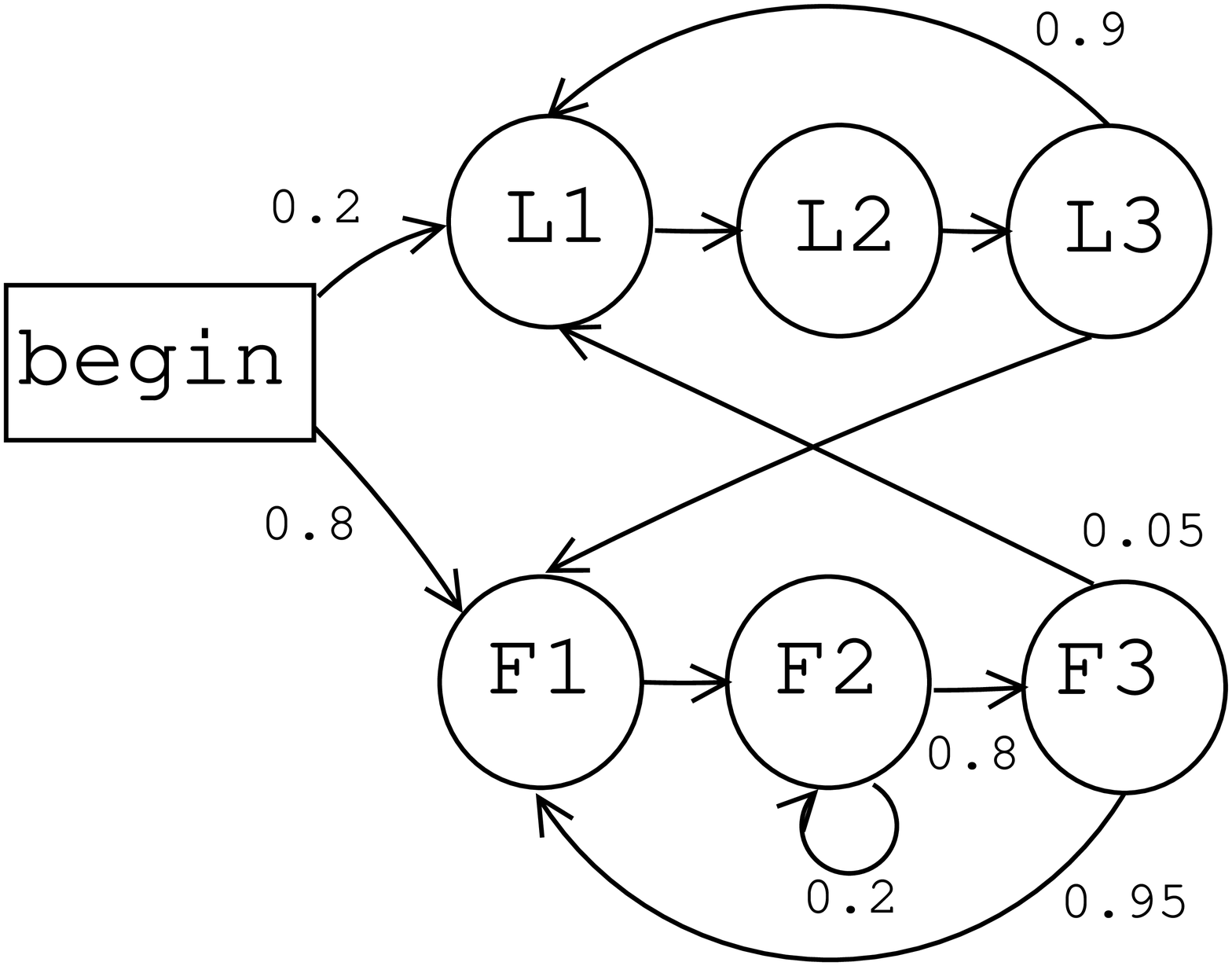}}}\par}
\caption{Occasionally dishonest casino (Model L3F3).
For the emission probabilities see Figure \ref{fig:lf}.}
\label{fig:l3f3}
\end{center}
\end{figure}

Two different types of data are used to score the posterior-Viterbi 
algorithm, namely synthetic and real data. In the former case, 
we start with the simple \textit{occasionally dishonest casino} illustrated 
in \cite{Durbinetal.1998}, referred here as \textit{LF} 
model (Figure \ref{fig:lf}); then we increase 
the complexity of the automaton with other two models. First, 
we introduce the \textit{occasionally dishonest casino} reported in 
Figure \ref{fig:l2f2} and referred as $L2F2$, in which fair (label F) 
and loaded dice (label L) come always in pairs (or one die is 
always tossed twice). A third more complex version of the \textit{occasionally 
dishonest casino} is shown in Figure \ref{fig:l3f3} (model $L3F3$).
In $L3F3$ the loaded dice are multiple of three (or one die is always tossed 
three times), while the number of fair tosses are at least three 
but can be more.

Accordingly, for each toy model presented above (Figures 
\ref{fig:lf}, \ref{fig:l2f2} and \ref{fig:l3f3}), 
we produced 50 sequences of 300 dice outcomes and we 
trained the corresponding empty models (one for each models) using the Baum-Welch 
algorithm. The initial empty models have the same topology of 
the models $LF$, $L2F2$ and $L3F3$, with their emission 
and allowed transition probabilities set to the uniform distribution.

After training, we tested the ability of different algorithms 
(Viterbi, 1-best and PV) to recover the original \textit{labeling} from 
the observed sequence of numbers (the dice outcomes). 

The problem of the prediction of the all-beta transmembrane regions is used 
to test the algorithm on real data application. In this case 
we use a set that includes 20 constitutive beta-barrel membrane 
proteins whose sequences are less than 25\% homologous and whose 
3D structure have been resolved. The number of beta-strands forming 
the transmembrane barrel ranges from 2 to 22. Among the 20 proteins 
15 were used to train a circular HMM (described in \cite{Martellietal.2002}), 
and here are tested in cross-validation (1a0sP, 1bxwA, 
1e54, 1ek9A, 1fcpA, 1fep, 1i78A, 1k24, 1kmoA, 1prn, 1qd5A, 1qj8A, 
2mprA, 2omf, 2por). Since there is no detectable sequence identity 
among the selected 15 proteins, we adopted a leave-one-out approach 
for training the HMM and testing it. All the reported results 
are obtained during the testing phase, and the complete set of 
results is available at www.biocomp.unibo.it/piero/posvit. 

The other 5 new proteins (1mm4, 1nqf, 1p4t, 1uyn, 1t16) are used 
as a blind new test.

\subsubsection*{Measures of accuracy}

We used three indices to score the accuracy of the algorithms. 
The first one is $Q_2$ which computes the number of correctly assigned 
labels divided by the total number of observed symbols.
Then we use the \textit{SOV} index \cite{Zemlaetal.1999} 
to evaluate the segment overlaps.
Finally, in the case of the all-beta transmembrane proteins we 
adopt a very stringent measure called $Q_{ok}$: a prediction is considered 
correct \textit{only if the number of transmembrane segments coincides 
with the observed one and the corresponding segments have a minimal 
overlap of m residues} \cite{Farisellietal.2003}. 
The value \textit{m} is segment-dependent  and for each segment pairs, 
is computed as 
\begin{eqnarray}
m= min \{ |seg_{pr}|/2,|seg_{ob}|/2 \} 
\end{eqnarray}
where $|seg_{pr}|$ and $|seg_{ob}|$ are the predicted and observed 
segment lengths, respectively.

\section*{Results and Discussion}

\subsubsection*{Testing the decoding algorithms on toy models}

We start using one of the simplest HMM model that can be thought 
of (\textit{LF}), which is the \textit{occasionally dishonest casino} presented 
in \cite{Durbinetal.1998}. 
\textit{LF} can parse any kind of observed sequence of numbers 
ranging from 1 to 6 (the die faces), generated with loaded and 
fair dice. Based on the \textit{LF} model we produced 50 sequences 
with 300 dice outcomes and we trained an empty model with them. 
After this, we tested the three decoding algorithms that preserve 
the automaton grammar on the task of reconstructing the correct 
labeling. 

In Table \ref{tab:toymodels}, we show that the accuracy of the posterior-Viterbi 
is greater than those of the other two algorithms. It is worth 
noticing that with this simple model the posterior algorithm 
alone achieves a similar accuracy (data not shown).
 
The $L2F2$ and $L3F3$ models, in which no one of the 
posterior decoding reconstructions is consistent with the automaton grammar 
(not parsable) are of some interest. In this case, among the 
three grammar-preserving algorithms, the posterior-Viterbi is 
the best performing one. This is particularly true for the $L3F3$
model, in which the SOV values highlight a quite good performance 
of PV. Considering that the three reconstructed models, as computed 
with the Baum-Welch, are very similar to the theoretical ones 
and independent from decoding, it is worth noticing the performance drop 
of the Viterbi algorithm. From these results it appears that in 
some cases the use of the PV decoding leads to a better performance 
given the same data and the same model.

\subsubsection*{Testing the decoding algorithms on real data}

In order to test our decoding algorithm on real biological data, 
we used a previously developed HMM, devised for the prediction 
of the topology of beta-barrel membrane proteins \cite{Martellietal.2002}. The hidden 
Markov model is a sequence-profile-based HMM and takes advantage 
of emitting vectors instead of symbols, as described in \cite{Martellietal.2002}.

Since the previously designed and trained HMM \cite{Martellietal.2002} emits profile 
vectors, sequence profiles have been computed from the alignments 
as derived with PSI-BLAST \cite{Altschuletal.1997} on the non-redundant database of 
protein sequences (ftp://ftp.ncbi.nlm.nih.gov/blast/db/) .

The results obtained using the four different decoding algorithms 
are shown in Table \ref{tab:intramodel}, 
where the performance is tested with a jack-knife 
validation procedure for the first 15 proteins and as blind-test 
for the latter 5 (see Methods). It is evident that for the problem 
at hand the Viterbi decoding and the 1-best are unreliable, since 
only one of the proteins is correctly assigned. In this case 
the posterior decoding is more efficient and can correctly assign 
60\% and 40\% of the proteins, in cross-validation and on the 
blind set, respectively. Here the posterior decoding is used 
without MaxSubSeq , introduced before to recast the grammar \cite{Martellietal.2002}. 

From Table \ref{tab:intramodel} 
it evident that the new PV decoding is the best performing 
decoding achieving 80\% and 60\% accuracy in cross-validation 
and on the blind set, respectively. This is done ensuring that predictions 
are consistent with the designed automaton grammar.

\subsubsection*{Comparison with other available HMMs}

Although this is out of the scope of this paper, the reader may 
be interested in seeing a comparison between our HMM-decoding 
with those obtained from the available web servers, based on 
similar approaches \cite{Bagosetal.2004, Bigelowetal.2004}. 
In Table \ref{tab:indexcomparison} we show the results. The \textbf{tmbb} 
server \cite{Bagosetal.2004} allows the user to test three different algorithms, 
namely Viterbi, 1-best and posterior. Differently from us they 
find that their HMM does not show significant differences among 
the three decoding algorithms. This dissimilar behaviour may 
be due to several concurring facts: \textit{i}) the different HMM 
models, \textit{ii}) \textbf{tmbb} runs on a single-sequence input, \textit{iii}) \textbf{tmbb} 
is trained using the \textit{Conditional Maximum Likelihood} \cite{Krogh1994}.
 
The second server \textbf{PROFtmb} \cite{Bigelowetal.2004} 
is based on a method that exploits 
multiple sequence information and posterior probabilities. Their 
decoding is related to the posterior-Viterbi; however, in their 
algorithm the authors first obtained the posterior sum contracted into 
two possible labeling (inner/outer loops and transmembrane as 
we did in \cite{Martellietal.2002}), 
then they made use of the explicit value of 
the HMM transition probabilities ($a_{i,j}$). In this way they count the 
transition probabilities twice (implicitly in the posterior-probability 
and directly into their algorithm) and the \textbf{PROFtmb} performance 
is not very different from ours. In our opinion, the fact that 
the newly implemented PV algorithm performs similarly or better, 
with respect to all indices, suggests that PV can be useful also when applied 
to the other HMM models.

\section*{Conclusions}

The new PV decoding algorithm is more convenient in that overcomes 
the difficulties of introducing a problem-dependent optimization 
algorithm when the automaton grammar is to be re-cast. When one-state-path 
dominates we may expect that PV does not perform better than 
the other decoding algorithms, and in these cases the 1-best 
is preferred \cite{Krogh1997}. Nevertheless, we show that when several 
concurring paths are present, as in the case of our beta-barrel 
HMM, PV performs better than the others.

A performance similar to that obtained with PV decoding can be 
achieved using MaxSubSeq algorithm \cite{Farisellietal.2003} on top of the posterior 
sum decoding. However, although MaxSubSeq is a very general two-class 
segment optimization algorithm, PV is far more useful when the 
underlying predictor is a HMM, where more than two labels and 
different constraints can be introduced into the automaton grammars. 
\\
Although PV takes a time longer than other algorithms (the posterior + the Viterbi 
time), the PV asymptotic computational time-complexity still 
remains, as for the other decodings $O(N^2\cdot L)$ 
(where \textit{L} and \textit{N} are 
the protein length and the number of states, respectively). As 
far as the memory requirement is concerned, PV needs the same 
space-complexity of the Viterbi and posterior ($O(N\cdot L)$),
while 1-\textit{best} in the average case requires less memory, and 
can also be reduced  \cite{Krogh1997}. 
When computational speed is an issue, 
Viterbi algorithm is the fastest and the time complexity order 
is $time(viterbi)\le time(1-best) \le time(PV)$.

Finally, PV satisfies any HMM grammar structures, including automata 
containing silent states, and it is applicable to all the possible 
HMM models with an arbitrary number of labels and without having 
to work out a problem-dependent optimization algorithm.

\section*{List of abbreviations}
\begin{itemize}
\item {\bf HMM}: hidden Markov model.
\item {\bf PV}: Posterior-Viterbi.
\end{itemize}

\section*{Authors' contributions}

PF developed the Posterior-Viterbi algorithm. PLM designed and 
trained the Hidden Markov Models. RC contributed to the problem. 
PF, PLM and RC authored the manuscript.

\section*{Acknowledgements}

This work was partially supported by the BioSapiens Network of 
Excellence, a grant of the Ministero della Universit\`{a} e della 
Ricerca Scientifica e Tecnologica (MURST), a grant for a target 
project in Biotechnology (CNR), a project on Molecular Genetics 
(CNR), a PRIN 2002 and a PNR 2001-2003 (FIRB art.8).


\newpage

\noindent 
\begin{table}
\caption{Accuracy of the different algorithms on the toy-models}
\begin{tabular}{lcccc}
\hline
Algorithms &     &   toy-models & \\
           &     LF  &     L2F2 &   L3F3 \\
\hline
viterbi & & & \\
Q2      &       0.80    &  0.86 &  0.47 \\
SOV     &       0.48    &  0.73 &  0.35  \\
SOV(L)  &       0.42    &  0.64 &  0.37  \\
        & & & \\
1-best & & & \\
Q2      &       0.80    &  0.86 &  0.88 \\
SOV     &       0.48    &  0.73 &  0.81  \\
SOV(L)  &       0.42    &  0.64 &  0.72  \\
        & & & \\
posterior-Viterbi & & & \\
Q2      &        0.82   &   0.88 &   0.90 \\
SOV     &        0.66   &   0.80 &   0.82 \\
SOV(L)  &        0.61   &   0.75 &   0.78 \\
\hline
\end{tabular}
\\
\small
For the indices see 'Measure of accuracy' section. 
SOV(L)= SOV computed for the loaded class only. 
\normalsize
\label{tab:toymodels}
\end{table}

\begin{table}
\caption{$Q_{ok}$ prediction accuracy obtained with the 
four different decoding algorithms on the real data}
\begin{tabular}{lccccc}
\hline
Proteins  &   Viterbi  &  1-best   &  posterior & posterior-Viterbi \\
\hline
         & & & & \\
{\em cross-validation}&&          &           &            \\
1a0spTOT &  -  &   -  & -   &   OK  \\ 
1bxwaTOT &  -  &   -  & OK  &   OK  \\
1e54     &  -  &   -  & OK  &   OK  \\
1ek9aTOT &  -  &   -  & OK  &   OK  \\
1fcpaTOT &  -  &   -  & -   &   -   \\
1fepTOT  &  -  &   -  & -   &   OK  \\
1i78a    &  -  &   -  & OK  &   OK  \\
1k24     &  -  &   -  & -   &   OK  \\
1kmoaTOT &  -  &   -  & OK  &   OK  \\
1prn     &  -  &   -  & -   &   -   \\
1qd5a    &  -  &   -  & OK  &   OK  \\
1qj8a    &  -  &   -  & OK  &   OK  \\
2mpra    &  -  &   -  & OK  &   OK  \\
2omf     &  -  &   -  & OK  &   OK  \\
2por     &  -  &   -  & -   &   -   \\
$<Q_{ok}>$&0.0 &  0.0  & 0.60 &  0.80  \\
         & & & & \\
{\em blind-test}&     &    &           &            \\
1mm4     &  -  &   -  & OK &   -   \\
1nqf     &  -  &   -  & -  &   OK  \\
1p4t     &  OK &  OK  & OK &   OK  \\
1uyn     &  -  &   -  & -  &   OK  \\
1t16     &  -  &   -  & -  &   -   \\
$<Q_{ok}>$&0.20& 0.20 & 0.40 &  0.60  \\
\hline
\end{tabular}
\\
$Q_{ok}>$: see Measures of Accuracy. 
\label{tab:intramodel}
\end{table}

\begin{table}
\caption{Posterior-Viterbi accuracy compared with other algorithms and
HMM models} 
\begin{tabular}{lccccc}
\hline
Method & Q2 & SOV & SOV(BetaTM) & SOV(Loop) & $Q_{ok}$ \\
\hline
{\em cross-validation$^4$}&   &  & &      &\\
Posterior-Viterbi$^1$& 0.82 & 0.87 & 0.92 & 0.81  & 0.80\\
Viterbi$^1$          & 0.63 & 0.33 & 0.27 & 0.35  & 0.0\\
1-best$^1$           & 0.65 & 0.37 & 0.31 & 0.38  & 0.0\\
PROFTmb$^2$          & 0.83 & 0.87 & 0.88 & 0.84  & 0.73 \\
tmbb$^3$ (Viterbi)   & 0.78 & 0.83 & 0.81 & 0.82  & 0.60\\
tmbb$^3$ (1-best)    & 0.78 & 0.83 & 0.81 & 0.82  & 0.60\\
tmbb$^3$ (posterior) & 0.78 & 0.82 & 0.80 & 0.82  & 0.60\\
         & & & & &\\
{\em blind-test$^4$}      &   &  & &       &\\
Posterior-Viterbi$^1$& 0.80 & 0.81 & 0.84 & 0.74  & 0.60\\
Viterbi$^1$          & 0.62 & 0.38 & 0.35 & 0.40  & 0.20\\
1-best$^1$           & 0.63 & 0.38 & 0.36 & 0.40  & 0.20\\
PROFTmb$^2$          & 0.72 & 0.65 & 0.72 & 0.58  & 0.40 \\
tmbb$^3$ (Viterbi)   & 0.71 & 0.73 & 0.79 & 0.71  & 0.20\\
tmbb$^3$ (1-best)    & 0.71 & 0.73 & 0.79 & 0.71  & 0.20\\
tmbb$^3$ (posterior) & 0.72 & 0.75 & 0.81 & 0.71  & 0.20\\
\hline
\end{tabular}
\\
$1$ Model taken from Martelli et al., 2002 \cite{Martellietal.2002}\\
$2$ Bigelow et al., (2004) \cite{Bigelowetal.2004} \\
$3$ Bagos et al., 2004 \cite{Bagosetal.2004} \\
$4$ this is only referred to posterior-Viterbi decoding \\
\label{tab:indexcomparison}
\end{table}

\newpage


\end{document}